\documentclass[aps,prd,superscriptaddress,showpacs,nofootinbib,eqsecnum,amsfonts,amsmath]{revtex4-1}
\usepackage{amssymb}
\usepackage{amsmath}
\usepackage[makeroom]{cancel}
\usepackage{slashed}

\def\nn{\nonumber}
\def\beq{\begin{equation}}
\def\eeq{\end{equation}}
\def\be{\begin{equation}}
\def\ee{\end{equation}}
\def\bea{\begin{eqnarray}}
\def\eea{\end{eqnarray}}

\begin{document}

\title{Perturbatively conserved higher nonlocal charges of free-surface deep-water gravity waves}

\author{Andr\'e Neveu}
\affiliation{Laboratoire Charles Coulomb (L2C), UMR 5221 CNRS-Universit\'{e} de Montpellier, 34095 Montpellier, France}

\begin{abstract}
We exhibit a set of six explicit higher nonlocal charges of free-surface deep-water gravity waves conserved in lowest 
nontrivial orders of perturbation in the amplitude of the surface displacement.  
\end{abstract}
\maketitle
\section{Introduction}
Inviscid irrotational deep-water gravity waves in one dimension are a much studied subject in
mathematical physics. Beyond its conceptual simplicity, a large part of its appeal comes from the fact that 
in some sense it comes close to integrability~\cite{Zakharov1994}. Here, by choosing a convenient set of
dynamical variables, we give explicit expressions for six nonlocal charges 
beyond energy and momentum, conserved in lowest orders of perturbation in powers of the vertical
displacement of the surface. In the linearized approximation
two of them generalize momentum and energy conservation with 
higher spatial derivatives. As far as we know, the other ones are unknown in this approximation.

In section II we show that the velocity potential~${\cal U}(x,z=0,t) \equiv U(x,t)$ at the rest altitude $z=0$
together with the surface position~$\eta(x,t)$ of the fluid are convenient variables to study the 
Euler equations~\cite{Euler1757} and their conservation laws in perturbation. With these fundamental variables we give
perturbative expansions of~${\cal U}(x,z,t)$ and its $z$ derivative at the surface where the 
harmonic function in $x$ and $z$ associated with~${\cal U}(x,z,t)$ is introduced and will play an important r{\^o}le

$${\cal V}(x,z,t)=-\frac{1}{\pi}\int_{-\infty}^{+\infty}\frac{x-x'}{(x-x')^2+z^2} U(x',t){\rm d}x'$$.

In section III  we give a few mathematical formulas involving several principal value integrals which are necessary 
to derive the conservation of the nonlocal charges.

In section IV we present these nonlocal charges, beginning with generalizations of energy and momentum,
and new ones using~${\cal V}(x,z=0,t) \equiv V(x,t)$ already in the lowest order, linear approximation,
of the equations of motion.

In the discussion, section V, we comment on the connection with the nonintegrability of the Euler equations and on the
analogy with the~$\varphi^4$ model in two dimensions.

\section{Perturbative expansion of the Euler equations for gravity waves}
The velocity field with horizontal component~$u(x,z,t)$  and vertical component~$v(x,z,t)$ of an 
incompressible inviscid fluid of unit mass per unit volume in the gravity field~$g$ satisfies 
the Euler equations of motion~\cite{Euler1757}

\bea
&u_t +uu_x+ vu_z =-p_x \nn \\
&v_t + uv_x + vv_z= -g-p_z
\label{Euler}
\eea
\noindent with $p$ the pressure field.

We consistently restrict ourselves to irrotational motions. In such a motion, the velocity 
field~$u(x,z,t)$,~$v(x,z,t)$ derives from a velocity potential~${\cal U}(x,z,t)$

$$u=\partial_x {\cal U}, \qquad  v=\partial_z {\cal U} $$

\noindent and~${\cal U}$ is a harmonic function of~$x$ and~$z$

$$\partial_x\, {\cal U}=\partial_z\, {\cal V}, \qquad  \partial_z\, {\cal U}=-\partial_x \,{\cal V}$$.

In an infinitely deep fluid, the boundary condition at~$z \rightarrow -\infty$ is~${\cal U} \rightarrow 0$
and at the surface 

$$v(x,\eta (x,t),t) - u(x,\eta (x,t),t) \eta (x,t)_x =\eta (x,t)_t$$. \\

The equations of motion can be derived from the Lagrangian

\bea
{\cal L}=\frac{1}{2}\int _{-\infty}^{+\infty}{\rm d}x\int _{-\infty}^{\eta}({\cal U}_x^2 + {\cal U}_z^2){\rm d}z
-\frac{1}{2}g \int _{-\infty}^{\infty}\eta^2{\rm d}x  
+ \int _{-\infty}^{\infty}{\rm d}x \lambda \bigl[\eta _t -{\cal U}_z\bigl(x,\eta (x)\bigr) +{\cal U}_x\bigl(x,\eta (x)\bigr)\eta _x \bigr]
\label{Lagrangian}
\eea

\noindent together with the boundary condition~${\cal U} \rightarrow 0$ at~$x \rightarrow \pm \infty$. $\lambda(x,t)$ is the
Lagrange multiplier which imposes the boundary condition at the surface.

One finds that~$\lambda(x,t)={\cal U}\bigl(x,\eta (x),t\bigr)$ satisfies these equations, which then write:

\bea
&\partial_x^2\, {\cal U} + \partial_z^2\, {\cal U}=0,  \nn \\
&\partial_t\lambda+{\cal U}_x\bigl(x,\eta (x)\bigr)\;\partial_x\lambda -\frac{1}{2}\Bigl[{\cal U}_x\bigl(x,\eta (x)\bigr)^2 
+{\cal U}_z\bigl(x,\eta (x)\bigr)^2\Bigr]
+g\eta=0.
\label{motion}
\eea

The harmonic function~${\cal U}(x,z,t)$ can be recovered for all~$z<0$ from its value at~$z=0$ which 
we call~$U(x,t)$ by the formula

\be 
{\cal U}(x,z,t)=-\frac{z}{\pi}\int _{-\infty}^{+\infty}\frac{U(x',t)}{(x-x')^2+z^2}{\rm d}x'
\label{continuation}
\ee

\noindent and its~$z$-derivative is then

$$-\frac{\partial \,{\cal U}(x,z)}{\partial z}=\frac{1}{\pi}\int _{-\infty}^{+\infty}U(x')
\frac{(x-x')^2-z^2}{\bigl((x-x')^2+z^2\bigr)^2}{\rm d}x'$$.

We also introduce the harmonic stream function~${\cal V}(x,z,t)$ by

$${\cal V}(x,z,t)=-\frac{1}{\pi}\int _{-\infty}^{+\infty}\frac{(x-x')}{(x-x')^2+z^2}U(x'){\rm d}x'$$

\noindent for~$z<0$ so that~${\cal U}+{\rm i}{\cal V}$ is a function of~$x+{\rm i}z$ and~${\cal V}(x,0,t)$ is given by 
a principal value integral which we call~$V(x,t)$. One has then

$$\frac{\partial\, {\cal U}(x,z)}{\partial x} =\frac{\partial \,{\cal V}(x,z)}{\partial z}\qquad
\frac{\partial\, {\cal U}(x,z)}{\partial z} =-\frac{\partial \,{\cal V}(x,z)}{\partial x}$$

\noindent and as~$z\rightarrow 0_-$, $\partial_z \,{\cal U}(x,z)$ is given by the principal value integral

$$\lim_{\epsilon\rightarrow 0}\frac{1}{\pi}\int _{-\infty}^{+\infty}\frac{(x-x')}{(x-x')^2+\epsilon^2}U(x')_{x'}{\rm d}x'
= -\frac{\partial \,V(x)}{\partial x}$$.

Equation~(\ref{continuation}), valid for~$z<0$, does not mean that~${\cal U}(x,z)$ is odd in~$z$, but around~$z=0$ it can be expanded
in powers of~$z$ as

\be
  {\cal U}(x,z)=U(x)- zV(x)_x
-\frac{1}{2}z^2 U(x)_{xx} +\frac{1}{6}z^3 V(x)_{xxx} + {\cal O}(z^4).
\label{expansion}
\ee

From this, one obtains the perturbative expansion of~$\lambda$ by setting~$z=\eta(x)$ in this equation.\\

The usual Hamiltonian, kinetic plus potential energies, is 

$${\cal H}=\frac{1}{2}\int _{-\infty}^{+\infty}{\rm d}x\int _{-\infty}^{\eta(x)}({\cal U}_x^2 + {\cal U}_z^2){\rm d}z
+\frac{1}{2}g \int _{-\infty}^{\infty}\eta^2{\rm d}x$$.

Using the equations of motion and the boundary conditions, one finds that this can be reduced 
to an integral over~$x$ only, to give the following
expression for the total energy of the motion in terms of the dynamical variables at the surface:

\be
E_1=\int _{-\infty}^{+\infty}\Bigl[\frac{1}{2}{\cal U}\bigl(x,\eta (x)\bigr){\cal U}_z\bigl(x,\eta (x)\bigr)
-\frac{1}{2}\,{\cal U}\bigl(x,\eta (x)\bigr){\cal U}_x\bigl(x,\eta (x)\bigr)\eta_x(x)
+\frac{1}{2}g\,\eta (x)^2\Bigr]{\rm d}x
\label{energy1}
\ee

\noindent and for the canonical total momentum we have

\be
P_1=\int _{-\infty}^{+\infty}\lambda(x)\eta(x)_x{\rm d}x=\int _{-\infty}^{+\infty}{\rm d}x\int _{-\infty}^{\eta(x)}{\cal U}_x {\rm d}z.
\label{momentum1}
\ee

In the linearized approximation, the equations of motion reduce to

\be
\lambda(x,t)_t=U(x,t)_t=-g\eta(x,t), \qquad
\eta(x,t)_t=\lim_{\epsilon\rightarrow 0}\frac{1}{\pi}\int _{-\infty}^{+\infty}\frac{(x-x')}{(x-x')^2+\epsilon^2}U(x')_{x'}{\rm d}x'
\ee

\noindent so that for a plane wave~$\exp({\rm i}(k x -\omega t))$ one recovers the usual dispersion law~$\omega^2=g|k|$.

\section{Mathematical Formulas}

When expanding in perturbation the equation of motion~({\ref{motion}}) for~$\lambda$ and the boundary condition
at the surface

\be
\eta(x,t)_t -{\cal U}_z\bigl(x,\eta (x),t\bigr) +{\cal U}_x\bigl(x,\eta (x),t\bigr)\eta(x,t)_x=0,
\label{boundary}
\ee

\noindent one encounters products of several principal value integrals which require the evaluation 
for example of

\be
\lim _{\epsilon_1\rightarrow 0,\epsilon_2\rightarrow 0}\int _{-\infty}^{+\infty}\frac{x-x'}{(x-x')^2+\epsilon_1^2}
\frac{x-x''}{(x-x'')^2+\epsilon_2^2}{\rm d}x.
\ee

This is straighforward by contour integration at infinity in the complex plane, and in the limit of the epsilons 
going to zero gives~$\pi^2 \delta(x'-x'')$. 

Here are other useful identities. For clarity we have
not mentioned explicitly that in these identities all the epsilons are independent from one another, and that they are 
just there to remind us that we are actually dealing in the end with principal value integrals.

\be
\frac{x-x'}{(x-x')^2+\epsilon^2}
\frac{x-x''}{(x-x'')^2+\epsilon^2}=\frac{x-x'}{(x-x')^2+\epsilon^2}
\frac{x'-x''}{(x'-x'')^2+\epsilon^2}+\frac{x-x''}{(x-x'')^2+\epsilon^2}
\frac{x''-x'}{(x''-x')^2+\epsilon^2}+\pi^2 \delta(x-x')\delta(x-x''),
\ee

\be
\frac{x-x'}{(x-x')^2+\epsilon^2}
\frac{x-x''}{(x-x'')^2+\epsilon^2}=\frac{1}{2}\frac{x-x'}{(x-x')^2+\epsilon^2}
\frac{x'-x''}{(x'-x'')^2+\epsilon^2}+\frac{\pi^2}{2} \delta(x-x')\delta(x'-x'')
\ee

\noindent when this second equation is multiplied by an expression  symmetric in~$x$ and~$x''$, and

\begin{multline}
\int _{-\infty}^{+\infty}\frac{x-x'}{(x-x')^2+\epsilon^2}
\frac{x-x''}{(x-x'')^2+\epsilon^2}\frac{x-x'''}{(x-x''')^2+\epsilon^2}{\rm d}x\\
=\pi^2\Big(\frac{x'-x''}{(x'-x'')^2+\epsilon^2}\delta(x'-x''')
+\frac{x''-x'''}{(x''-x''')^2+\epsilon^2}\delta(x''-x')+\frac{x'''-x'}{(x'''-x')^2+\epsilon^2}\delta(x'''-x'')\Bigr).
\end{multline}

\section{New conserved quantities}

In many known integrable one-space-one-time dynamical systems, the higher conserved quantities 
appear in the weak field limit as bilinear expressions involving higher derivatives of the dynamical
variables which generalize energy and momentum densities. For deep-water waves it is first natural
to start with such generalizations. In lowest order the first generalization of momentum conservation would thus be

\be
P_2=\int _{-\infty}^{+\infty}\lambda(x)_x\eta(x)_{xx}{\rm d}x
\label{momentum2.0}
\ee

\noindent and the first generalization of energy conservation would be 

\be
E_2=\int _{-\infty}^{+\infty}\Bigl[\frac{1}{2}{\cal U}\bigl(x,\eta(x)\bigr)_x{\cal U}_{xz}\bigl(x,\eta(x)\bigr)
+\frac{1}{2}g\,{\eta_x(x)}^2\Bigr]{\rm d}x,
\label{energy2.0}
\ee

We shall come back later in this section to the extension of these two quantities to higher orders, and first
explore another route towards new conserved quantities. \\

The canonical energy~(\ref{energy1}) involves~${\cal U}_z(x,0)$ which is given
by a principal value integral in terms of the variables~$\lambda$ and~$\eta$. So, it is natural to look 
for other quantities which would be conserved already in lowest order and would similarly involve
nonlocal expressions in terms of~$\lambda$ and~$\eta$.

One sees immediately that in the weak field limit

\be
\int _{-\infty}^{+\infty}V(x)\eta(x){\rm d}x
\label{Veta}
\ee

\noindent is time independent. In next order, considering the equation of motion~(\ref{motion}) for~$\lambda$
and the expansion in powers of~$z$~(\ref{expansion}), it is more convenient
to replace~$U(x)$ by~$\lambda$ in the definition of~$V(x)$. 
In the remainder of this section, we adopt this new starting point:

\be
V(x)=-\lim_{\epsilon\rightarrow 0}\frac{1}{\pi}\int_{-\infty}^{+\infty}\frac{x-x'}{(x-x')^2+\epsilon^2} \lambda(x'){\rm d}x'.
\label{newV}
\ee

One must use the identities of the previous section to reduce the cubic terms of the time derivative 
of~(\ref{Veta}) with this new definition of~$V$, and one finds that they can be cancelled by a rather simple cubic addition:

\be
\frac{{\rm d}}{{\rm d}t}\int _{-\infty}^{+\infty}\Bigl[V(x)\eta(x) +\frac{1}{2}\eta(x)^2\lambda(x)_x\Bigr]{\rm d}x
\ee

\noindent vanishes at third  order in the surface variables.\\

We have pushed the calculation to next order, where it becomes much more involved, and found that 

\begin{multline}
\frac{{\rm d}}{{\rm d}t}\int _{-\infty}^{+\infty}\Bigl[V(x)\eta(x) +\frac{1}{2}\eta(x)^2\lambda(x)_x
-\frac{1}{2\pi}\int _{-\infty}^{+\infty}\frac{(x-x')^2-\epsilon^2}{\bigl((x-x')^2+\epsilon^2\bigr)^2}
\eta(x')^2\eta(x)U(x)_x{\rm d}x'\\
-\frac{1}{2\pi}\eta(x)^2\int _{-\infty}^{+\infty}{\rm d}x'
\frac{x-x'}{(x-x')^2+\epsilon^2}\bigl(\eta(x')_{x'}U(x)_{x}-\eta(x)_{x}U(x')_{x'}\bigr)\Bigr]{\rm d}x
\end{multline}

\noindent vanishes at fourth  order. In this expression, the limit~$\epsilon \rightarrow 0$ is of course understood.\\

As the order increases, the number of derivatives increases, and of course also the number of ways to distribute 
them among the various perturbative quantities, beyond what seems tractable by hand.\\

Considering the relative simplicity of this new conserved quantity~(\ref{Veta}), we try starting points with 
higher derivatives in it, analogous to the generalizations of energy~(\ref{energy2.0}) and momentum~(\ref{momentum2.0}):

\be
\int _{-\infty}^{+\infty}V(x)_x\eta(x)_x{\rm d}x.
\label{Veta2.0}
\ee

For this quantity we have found that at third order in the dynamical variables

\begin{multline}
\frac{{\rm d}}{{\rm d}t}\int _{-\infty}^{+\infty}{\rm d}x\Bigl[V(x)_x\eta(x)_x -\frac{1}{2}\eta(x)^2\lambda(x)_{xxx}
-\frac{1}{\pi}\int _{-\infty}^{+\infty}\frac{(x-x')}{(x-x')^2+\epsilon^2}
\eta(x)\eta(x')_{x'x'}V(x)_x{\rm d}x'\\
-\frac{2}{\pi}\int _{-\infty}^{+\infty}{\rm d}x'\eta(x)_xV(x)_x
\frac{x-x'}{(x-x')^2+\epsilon^2}\eta(x')_{x'}\Bigr]
\end{multline}

\noindent vanishes.\\

We have even managed to treat a case with two more derivatives:

\be
\int _{-\infty}^{+\infty}V(x)_{xx}\eta(x)_{xx}{\rm d}x.
\label{Veta2.1}
\ee

In the time derivative of this expression, we found that at third order four different terms are involved,
which we list: 

\be
\eta(x)V_x(x)U_{5x}(x), \quad  \eta(x)_xV_x(x)U_{4x}(x),\quad   \eta(x)_{xx}V_x(x)U_{xxx}(x),\quad  
\eta(x)_{xxx}V_x(x)U_{xx}(x).
\label{Veta2.2}
\ee

An expression whose time derivative could cancel these terms is found to be of the form

\begin{multline}
A\,\eta(x)^2U(x)_{5x} + B\,\eta(x)_x^2U(x)_{xxx} +C\,\eta(x)_{xx}^2U(x)_{x} +D\,U(x)_{xxx}U(x)_{xx}V_x(x) +E\,U(x)_{4x}U(x)_{x}V_x(x)\\
+F\,\eta(x)V_x(x)\frac{(x-x')}{(x-x')^2+\epsilon^2}\eta(x')_{4x'} +G\,\eta(x)_xV_x(x)\frac{(x-x')}{(x-x')^2+\epsilon^2}\eta(x')_{x'x'x'}\\
+H\,\eta(x)_{xx}V_x(x)\frac{(x-x')}{(x-x')^2+\epsilon^2}\eta(x')_{x'x'} + J\,\eta(x)_{xxx}V_x(x)\frac{(x-x')}{(x-x')^2+\epsilon^2}\eta(x')_{x'}.
\label{Veta2.3}
\end{multline}

The time derivative of this involves five more terms:

\begin{multline}
 \eta(x)_{4x}V_x(x)U_{x}(x), \;\;  \eta(x)_{xx}^2\eta(x)_x, \;\; 
\eta(x')_{4x'}\frac{(x-x')}{(x-x')^2+\epsilon^2}U(x)_x^2, \\  \eta(x')_{x'x'}\frac{(x-x')}{(x-x')^2+\epsilon^2}U(x)_{xx}^2, \;\; 
\eta(x)_{x}\eta(x'')_{x''x''}\eta(x')_{x'x'}\frac{(x-x')}{(x-x')^2+\epsilon^2}
\frac{(x-x'')}{(x-x'')^2+\epsilon'^2}.
\label{Veta2.4}
\end{multline}

So, to have conservation at third order, we arrive at a system of nine equations in nine unknowns, which has a unique solution:

$$A=\frac{1}{2},\; B=1,\; C=1,\; D=-\frac{6}{g},\; E=\frac{3}{g},\; F=-\frac{1}{\pi},\; G=-\frac{5}{\pi},\; 
H=-\frac{15}{\pi},\; J=-\frac{9}{\pi}$$.

Except for the coefficient~$A$, we see no obvious pattern in this increasing complexity when comparing with the cases
with fewer derivatives~(\ref{Veta}) and~(\ref{Veta2.0}).\\

There is still another relatively simple candidate for a new conservation law: The starting point~(\ref{Veta}) which is bilinear 
in~$U$ and~$\eta$ looks like the momentum~(\ref{momentum1}) also bilinear in~$U$ and~$\eta$. It is natural then to investigate
the conservation of a quantity which would resemble energy~(\ref{energy1}). Indeed, one finds immediately that to leading,
bilinear, order 

\be
\int _{-\infty}^{+\infty}{\rm d}x\Bigl[\frac{1}{2}\lambda_x(x)^2 +
\frac{g}{2\pi}\int _{-\infty}^{+\infty}{\rm d}x'\eta(x)\frac{(x-x')}{(x-x')^2+\epsilon^2}\eta_{x'}(x')\Bigr]
\label{U2+eta2}
\ee

\noindent is conserved.

In next, trilinear order, the time derivative of this expression gives the relatively simple cubic terms

\be
-\int _{-\infty}^{+\infty}{\rm d}x\Bigl[-\frac{1}{2}V_x(x)^2\lambda(x)_{xx} - g\eta(x)\eta_{xx}(x)V_x(x) +
\frac{g}{\pi}\int _{-\infty}^{+\infty}{\rm d}x'\eta(x)U(x)_x\frac{(x-x')}{(x-x')^2+\epsilon^2}\eta_{x'}(x')\Bigr].
\ee

To cancel this, one finds rapidly that one must add to~(\ref{U2+eta2}) no less than five terms

\begin{multline}
\int _{-\infty}^{+\infty}{\rm d}x\Bigl[A\,\eta(x)_x^2\eta(x) +B\,\eta(x)_xU(x)_xV(x)_x +C\,\eta(x)U(x)_{xx}V(x)_x \\ 
+D\,\int _{-\infty}^{+\infty}\eta_x(x)\frac{(x-x')}{(x-x')^2+\epsilon^2}
\eta_{x'}(x')\frac{(x'-x'')}{(x'-x'')^2+\epsilon^2}\eta(x''){\rm d}x'{\rm d}x''
+E\,\int _{-\infty}^{+\infty}\eta_x(x)\frac{(x-x')}{(x-x')^2+\epsilon^2}\eta_{x'}(x')U(x')_{x'}^2{\rm d}x'\Bigr].
\end{multline}

One obtains five equations in five unknowns, whose unique solution is 

$$A=0, \quad B=-1, \quad C=0, \quad D=-\frac{g}{\pi^2}, \quad E=-\frac{1}{2\pi}. $$\\

Let us now come to the generalizations to higher derivatives of energy~$E_2$~(\ref{energy2.0}) and momentum~$P_2$~(\ref{momentum2.0}).\\

The time derivative of~$E_2$ involves seven terms at trilinear order:

\begin{multline}
 V_{xx}U_xU_{xx}, \quad V_{x}U_{xx}^2, \quad \eta_x^2U_{xx}, \quad \eta^2U_{4x}, \quad 
V_x(x)\eta(x)\frac{(x-x')}{(x-x')^2+\epsilon^2}\eta(x')_{x'x'x'}, \\ V_x(x)\eta(x)_x\frac{(x-x')}{(x-x')^2+\epsilon^2}\eta(x')_{x'x'},
\quad V_x(x)\eta(x)_{xx}\frac{(x-x')}{(x-x')^2+\epsilon^2}\eta(x')_{x'}.
\end{multline}  \\

There are precisely seven terms whose time derivatives can cancel these:

\begin{multline}
A\,\eta_{xx}U_x^2 \; + B\,\eta U_{xx}^2 \; + C\,\eta_{xx}V_x^2 \; + D\,\eta V_{xx}^2 \; 
+ E\,\eta(x)^2\frac{(x-x')}{(x-x')^2+\epsilon^2}\eta(x')_{x'x'x'} \\ + F\,\eta(x)_x^2\frac{(x-x')}{(x-x')^2+\epsilon^2}\eta(x')_{x'} \;
+ G\,V_x(x)U_{xx}(x)\frac{(x-x')}{(x-x')^2+\epsilon^2}\eta(x')_{x'},
\end{multline}

\noindent and the cancellation occurs for the following values:

$$A=-\frac{1}{8}, \quad B=\frac{5}{8}, \quad C=-\frac{3}{8}, \quad D=\frac{1}{8}, \quad E=-\frac{g}{8\pi}, \quad F=-\frac{19g}{8\pi},
\quad G=\frac{3}{2\pi}.$$\\

The conservation of~$P_2$~(\ref{momentum2.0}) at third order is relatively simple. One finds:

$$\partial_t(\lambda_x\eta_{xx}) = -\frac{3}{2}\eta_xV_{xx}^2 + \frac{3}{2}\eta_xU_{xx}^2 $$

\noindent (up to total derivatives, of course) and we have found that this can be cancelled by adding just two trilinear terms:

$$\partial_t\Bigl[\lambda_x\eta_{xx}-\frac{3}{\pi}\eta_x(x)U_{xx}\frac{(x-x')}{(x-x')^2+\epsilon^2}\eta(x')_{x'}
-\frac{3}{2}\eta_x^2V_{xx}\Bigr]$$

\noindent vanishes at third order up to total derivatives.

In next order, for~$E_2$ as well as for~$P_2$ there are up to seven derivatives to be shared between~$\eta$,~$V$,~$U$ raised to
various powers, not to mention one or two principal values, and it seems that some formal manipulation software would 
be needed to arrive at a reliable conclusion.

\section{Discussion}

Craig and Worfolk~\cite{Craig-Worfolk1995}, and Dyachenko, Lvov and Zakharov~\cite{Zakharov1995} had given rather convincing evidence
that at fifth order in the surface displacement the Euler equations~(\ref{Euler}) are not integrable. 
However, a more recent publication~\cite{Dyachenko2019} has revived the conjecture of integrability by studying
the analytic structure, poles and cuts, of the complex velocity potential in the upper half-plane above the free surface,
using a new non-canonical Hamiltonian structure of the Euler equations~\cite{Lushnikov2019}. In this publication, 
new conserved quantities are found. How they could relate to those presented here goes beyond the scope of this
paper. An argument for or against integrability could come out of a fifth-order calculation of our six new quantities 
which could show whether they can or cannot be made time-independent.\\

An analogy can be made with the relativistic~$\varphi^4$ theory in 1+1 dimensions

\be 
\varphi_{tt} - \varphi_{xx} + \varphi - \frac{1}{6}\varphi^3=0
\label{phi^4}
\ee

\noindent as a low-amplitude, not integrable~\cite{Kruskal}, approximation  of the integrable sine-Gordon equation

\be
\varphi_{tt} - \varphi_{xx} + \sin\varphi=0 .
\label{sine-Gordon}
\ee

In both systems the total momentum is conserved of course:

$$\frac{{\rm d}}{{\rm d}t}\int _{-\infty}^{+\infty}\varphi_x \varphi_t \,{\rm d}x= 0,$$

\noindent and when trying a generalization with more derivatives as in our equation~(\ref{momentum2.0}), one finds that

$$\frac{{\rm d}}{{\rm d}t}\int _{-\infty}^{+\infty}\Bigl[\varphi_{xx} \varphi_{xt}+ \frac{1}{8}\bigl(\varphi^3 \varphi_{xt}
-\varphi_t^3 \varphi_{x}- \varphi_x^3 \varphi_{t}\bigr)\Bigr] {\rm d}x  
$$

\noindent vanishes in this quartic approximation in~$\varphi$, but not beyond, where the higher-order terms
of the expansion of the sine are necessary to achieve conservation. These approximate conservation laws could explain 
the long persistence of coherent structures in deep-water waves in our case and why approximate breathers, weak solutions 
of~(\ref{phi^4}), can be relevant in condensed matter physics~\cite{Campbell}.

\acknowledgments
We are grateful to Professor Zakharov for providing us with the recent references~\cite{Dyachenko2019} 
and~\cite{Lushnikov2019}.

\end{document}